\documentclass[aip,amsmath,amssymb,reprint]{revtex4-1}
\usepackage[dvipdfmx]{graphicx}
\usepackage{dcolumn}
\usepackage{bm}
\usepackage{color}
\usepackage{amsmath,cases}

\begin{document}
\title{Snell's law for spin waves at a 90-degree domain wall}

\author{Tomosato Hioki}
\affiliation{Institute for Materials Research, Tohoku University, Sendai 980-8577, Japan}

\author{Rei Tsuboi}
\affiliation{Institute for Materials Research, Tohoku University, Sendai 980-8577, Japan}

\author{Tom H. Johansen}
\affiliation{Department of Physics, University of Oslo, 0316 Oslo, Norway}
\affiliation{Institute for Superconducting and Electronic Materials, University of Wollongong, Northfields Avenue, Wollongong, NSW 2522, Australia}

\author{Yusuke Hashimoto}
\affiliation{Advanced Institute for Materials Research, Tohoku University, Sendai 980-8577, Japan}

\author{Eiji Saitoh}
\affiliation{Advanced Institute for Materials Research, Tohoku University, Sendai 980-8577, Japan}
\affiliation{Institute for Materials Research, Tohoku University, Sendai 980-8577, Japan}
\affiliation{Department of Applred Physics, Faculty of Engineering, The University of Tokyo, Tokyo 113-8656, Japan}

\date{\today}

\begin{abstract}
We report experimental observation of the refraction and reflection of propagating magnetostatic spin waves crossing a 90-degree domain wall (DW) in terms of time-resolved magneto-optical imaging. 
Due to the magnetization rotation across the 90-degree DW, the dispersion relation of magnetostatic spin waves rotates by 90 degrees, which results in the change in the propagation dynamics of spin waves in both sides of the DW.
We observe the refraction and reflection of magnetosatatc spin waves at the 90-degree DW, and reveal their characteristics, such as negative refraction.
The incident-angle dependence of the refraction angle is explained by the wavenumber conservation along the DW, same as the case of Snell's law for a light.
\end{abstract}

\maketitle

The relation of the angles between incident and refracted waves is called Snell's law.
In the field of magnonics, where the propagation of spin waves plays a main role for data processing and information transfer, the reflection and refraction of spin waves are important because it enables the manupilation of their phase and propagation orientation \cite{1,2,3,4,5,6,7,8,9,10,11}.
Refraction occurs due to the difference in the phase velocity between two different media, which is characterized by dispersion relation.
Spin waves in the small wavenumber regime (magnetostatic spin waves) have an anistoropic dispersion relation due to the anisotropic nature of magnetic dipole interaction. 
Therefore, the phase velocity of magnetostatic waves depends on the angle $\theta_\mathrm{\bf k}$ between the magnetization $\mathrm{\bf M}$ and the wavevector $\mathrm{\bf k}$.

The interplay between a magnetic domain wall (DW) and spin waves has been investigated, expanding the use of a DW for the various purposes in magnonics, such as a channel for spin waves and magnonic crystals\cite{12,13,14,15,16,17}.
The width of a DW, typically in the order of lattice spacing, is less than the wavelength of magnetostatic spin waves which is sub- to several micrometers in thin films\cite{SWlength}.
Therefore a DW acts for spin waves as an abruput magnetic boundary.

We report the refraction and reflection of magnetostatic spin waves by a 90-degree domain wall (DW) in a garnet film.
At a 90-degree domain wall, dispersion relation of magnetostatic spin waves rotates by 90 degrees due to the rotation of magnetization.
The direct observation of the magenetostatic spin waves by time-resolved magneto-optical imaging and the subsequent analysis of the observed images reveals the refraction law of magnetostatic spin waves at a 90-degree DW.
In ordinal refraction between two different media with positive refraction indices, the refracted wave has a wavevector directing forward along the boundary.  
Above certain incident angle, we observed that the refracted spin waves have a wavevector directing backwards along the DW.
This behavior is simlar to negative refraction, which refers to the refraction between media having diffrent refraction indices with opposite signs\cite{Ng1,Ng2}. 


\begin{figure}[t]
\includegraphics[width=8cm]{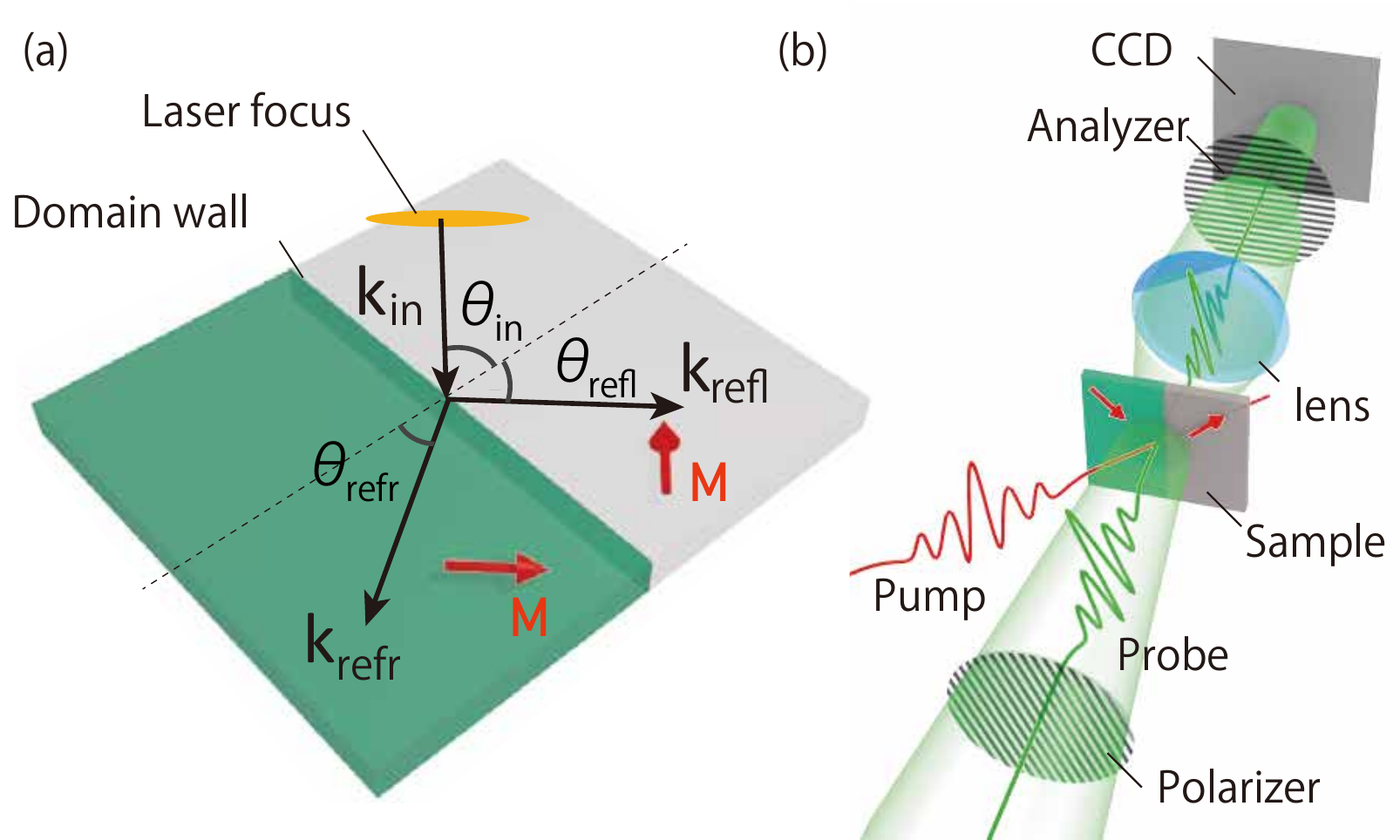}
\caption{\label{Fig1} (a) Schematic illustartion of sample configuration. Spin waves are excited by illuminating the sample by the focused pump beam near a 90-degree domain wall. The angle of the incident, reflected, and refracted spin waves by the domain wall are referred to $\theta_\mathrm{inc}$, $\theta_\mathrm{refr}$, and $\theta_\mathrm{refl}$, respectively.  (b) Schematic illustration of the time-resolved magneto-optical imaging. Faraday rotation angle of the transmitted probe beam is measured with the rotation analyzer method combined with a CCD camera.
}
\end{figure}

We use a Bi-doped garnet film with the composition of $\mathrm{Bi}_{0.7}\mathrm{Lu}_{2.3}\mathrm{Fe}_{4.2}\mathrm{Ga}_{0.8}\mathrm{O}_{12}$ (LuIG).
This sample is a  ferrimagnetic insulator having an in-plane spontaneous magnetization (4$\pi M_s$=780 G) due to the negative uniaxial anisotropy along the direction normal to the sample surface ($K_\mathrm{u}=-1.2\times 10^4$ erg cm$^{-3}$)\cite{18}.
In the absence of the external magnetic field, a Neel-type 90-degree domain wall (DW) is formed.
The DW can be pinned by the crystallographic defects in the sample.
The formation of magnetic domains has been confirmed by using Cotton-Mouton effect (CME)\cite{19}.
The width of the DW $\delta$ is calculated as 400 nm by $\delta = \pi \sqrt{A/K_\mathrm{c}}$, where $A$ ($= 3.7$ pJ/m) is the exchange coupling constant and $K_\mathrm{c}$ ($= 2.3 \times 10^2$ J/m$^3$) is the cubic anisotropy of the sample.
In our experiments, spin waves are excited by illuminating the sample with a 800 nm laser pulse (pump beam).
The excitation and propagation dynamics of spin waves are observed with a time-resolved magneto-optical imaging system based on the pump-and-probe technique and a rotation analyzer method using a CCD camera\cite{20}.
The spatial resolution of the image is 1 $\mu$m.
This setup measures the Faraday rotation angle of the transmitted pulse laser with the wavelength of 630 nm (probe beam).
By taking the difference between the image taken with and without the pump beam illumination, we observe the spatiotemporal magnetization change along the sample depth direction due to the pump beam.
The observed spin waves are analyzed with a model based on Fourier transform, called the spin-wave tomography (SWaT)\cite{18}.

In order to investigate the interplay between a DW and propagating spin waves, we used planar propagating spin waves generated by the illumination of a pump beam.
In our experiments, spin waves are generated by the optically-excited elastic waves through magnetoelastic coupling (MEC)\cite{Kittel, Ogawa}.
The amplitude of spin waves excited via MEC is resonantly enhanced  at the crossing of the dispersion curves of spin waves and elastic waves. 
This determines the wavenumber and frequency of the dominant spin wave propagation\cite{21}.
In the experiment, we chose the wavevector of elastic waves by using a slit, making a focus of the pump beam elliptical.
As a result, almost planar spin waves with very narrow $\bf k$ and $\omega$ distribution are obtained.

\begin{figure}[t]
\includegraphics[width=8cm]{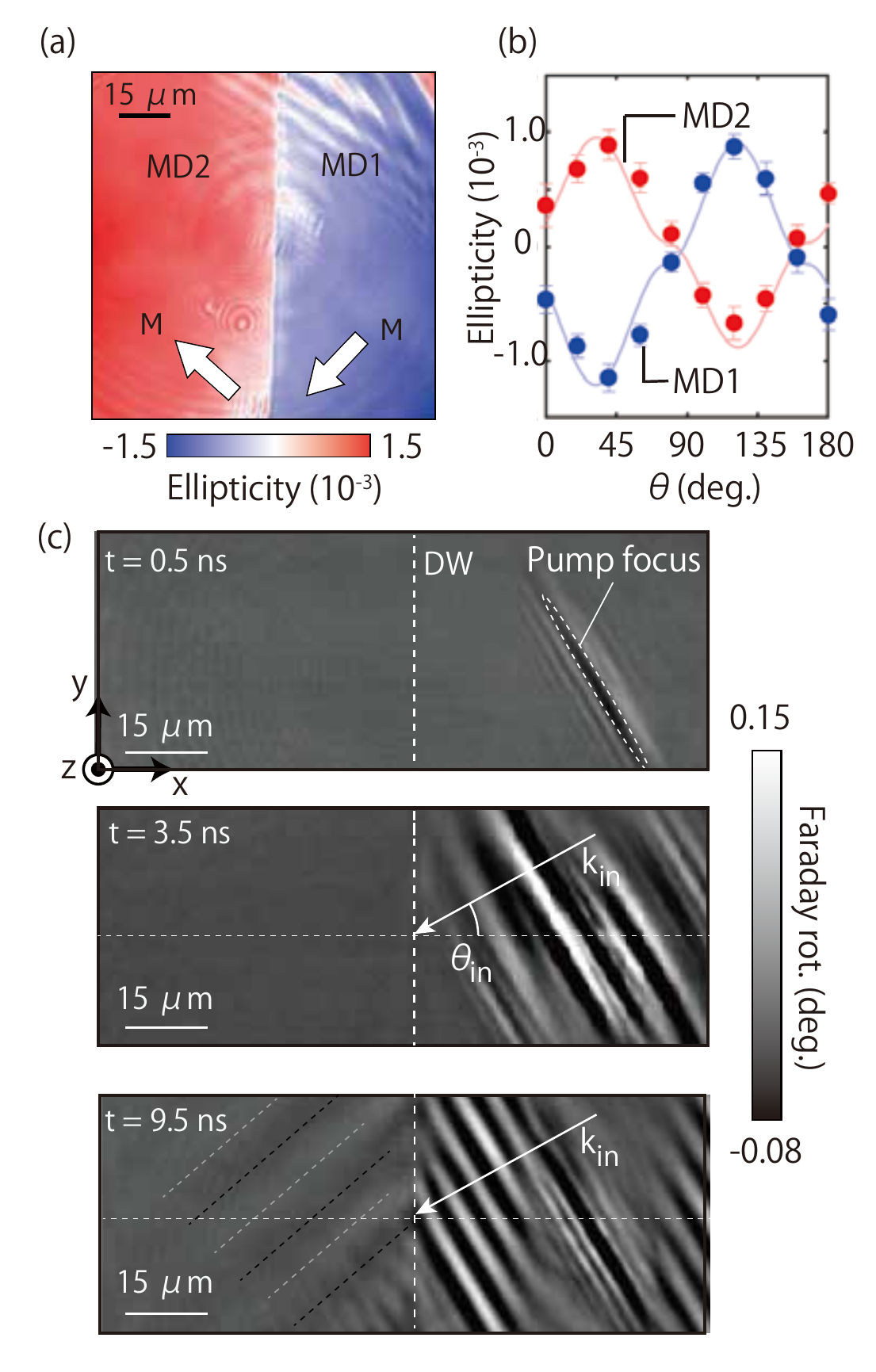}
\caption{\label{Fig2} (a) Static magneto-optical image obtained through the Cotton-Mouton effect which reflects the in-plane orientation of the magnetization. (b) The light ellipticity of the transmitted probe beam, reflecting the in-plane magnetization through the Cotton-Mouton effect, as a function of its polarization angle. The blue and red filled circles are obtained in MD1 and MD2, respectively. Solid line is the calculated polarization dependence for each domains\cite{19}. (c) Magneto-optical images obtained at different time delay between pump beam and probe beam. White arrows in the panel indicate the wavevector of the incident spin waves. The white dotted line indicates the position of the DW. 
}
\end{figure}

Let us first show a 90-degree DW observed by Cotton-Mouton effect (CME) in Fig.~\ref{Fig2}(a) .
CME refers to a magnetic birefringence effect which induces light ellipticity to the transmitted light. 
Two magnetic domains (MDs) separated by a DW are clearly observed.
We here name these domains MD1 and MD2 as defined in Fig.~\ref{Fig2}(a). 
The orientation of magnetization is determined by CME observed as a function of the polarization angle of the incident probe beam [Fig.~\ref{Fig2}(b)].
The light ellipticity obtained in MD1 and MD2 shows almost the same magnitude with opposite signs, meaning the magnetization orientations in MD1 and MD2 are orthogonal to each other.

Next, we show the spin wave propagation across the DW.
The spin waves are excited by illuminating the sample with a pump beam in MD1.
The propagation dynamics of spin waves are shown in Fig.~\ref{Fig2}(c).
In MD1, the spin waves propagate with the wave vector $\mathrm{\bf k}_\mathrm{in}$ as shown in the middle panel in Fig.~\ref{Fig2}(c).
Propagation of spin waves is consistent with that observed in a single domain sample.
We found the spin wave propagates across the DW and appears in MD2.
Interestingly, the wavevector of the spin waves in MD2 is opposite to what is expected for the ordinal refraction such as the refraction of a light at the interface between different media with positive refraction indices.
Namely the spin waves in MD2 represent the negative refraction of spin waves by the DW.

\begin{figure}[t]
\includegraphics[width=8.5cm]{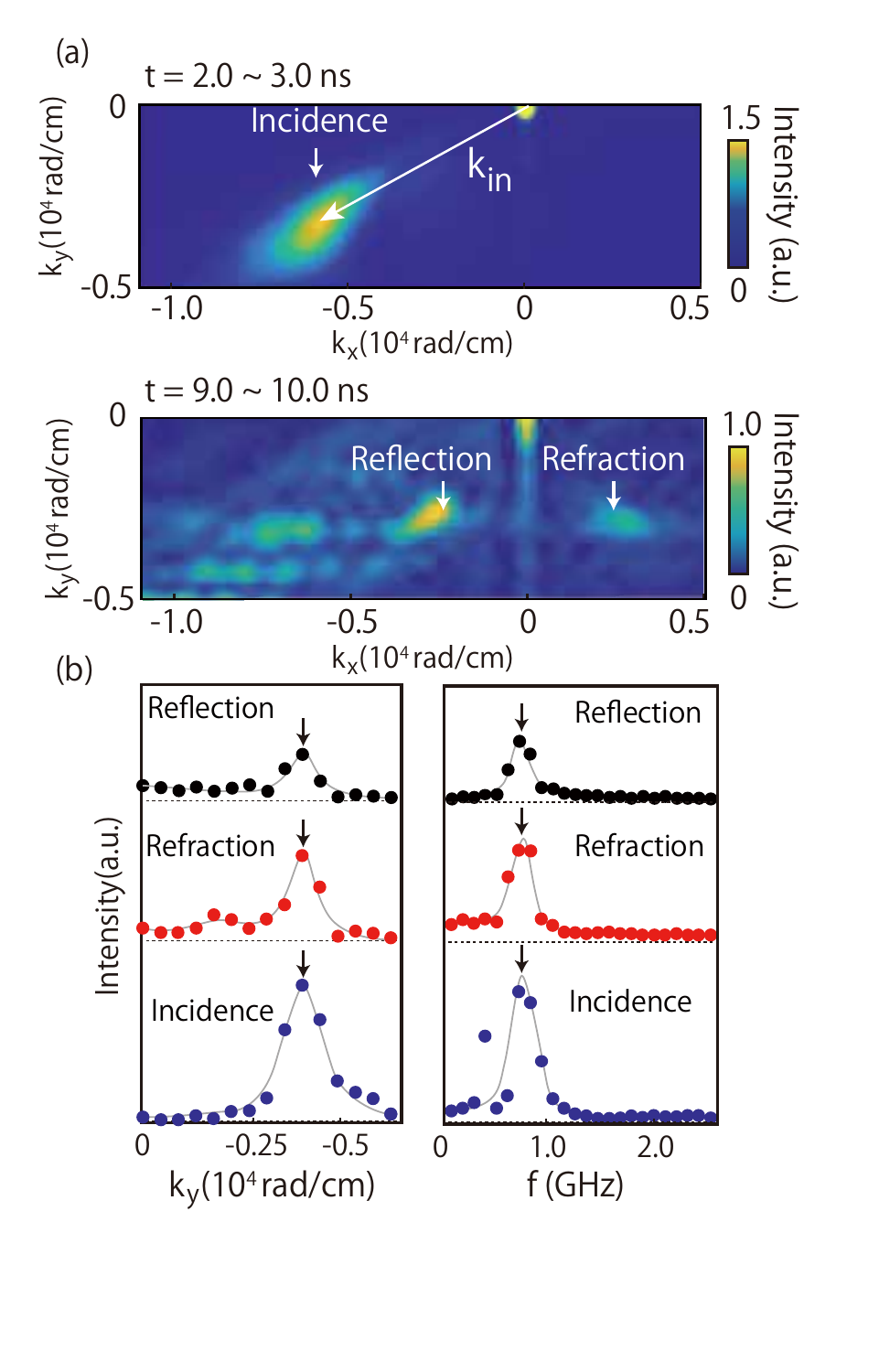}
\caption{\label{Fig3}(a) the SWaT spectra at $f=$ 1.0 GHz obtained by filtering the time range of $2.0$ ns to $3.0$ ns (top panel) and $9.0$ ns to $10.0$ ns (bottom panel). The single peak of the incident spin waves are split into reflected and refracted spin waves after the incidence onto the DW. (b) The left panel shows the integration of the SWaT spactra with respect to $x$ component of wavevectors. Integration around the peak representing incident, refracted, and reflected spin waves are shown by filled blue, red and black circles. The right panel shows the integrated the SWaT spectra over $\mathrm{\bf k}$ for different frequencies. The quadrant including the peak of incident, refracted, and reflected spin waves are used to obtain each specta. The gray solid line is the eyeguide in the both panels.
}
\end{figure}

The change in the $\bf k$ of the spin waves at the DW is clearly seen also in the SWaT spectra.
In order to distinguish spin waves propagating in MD1 and MD2, we applied a time window, given by a Gaussian function with the central time at $t_c$ and the width of 1.0 ns, to the calculation of the SWaT spectra.
Figure~\ref{Fig3}(a) shows a cross section of the obtained SWaT spectra at $\omega $ =1.0 GHz with $t_c =$2.5 ns, and 9.5 ns, respectively.
In the top panel of Fig.~\ref{Fig3}(a), we see a single strong peak reflecting the spin waves excited by the pump beam.
On the other hand, we see, in the bottom panel of Fig.~\ref{Fig3}(a), two peaks at different {\bf k} values. 
These two peaks are attributed to spin waves reflected and refracted by the 90-degree DW.
In order to confirm the conservation of wavevector along the DW ($k_y$) and frequency among the incident, refracted, and reflected waves, we show the integrated spectral intensity in Fig.~\ref{Fig3}(c).
We see that the $k_y$ and the central frequency of these three waves are the same, satisfying the presupposition to derive Snell's law for spin waves at a 90-degree domain wall. 

\begin{figure}[t]
\includegraphics[width=8.5cm]{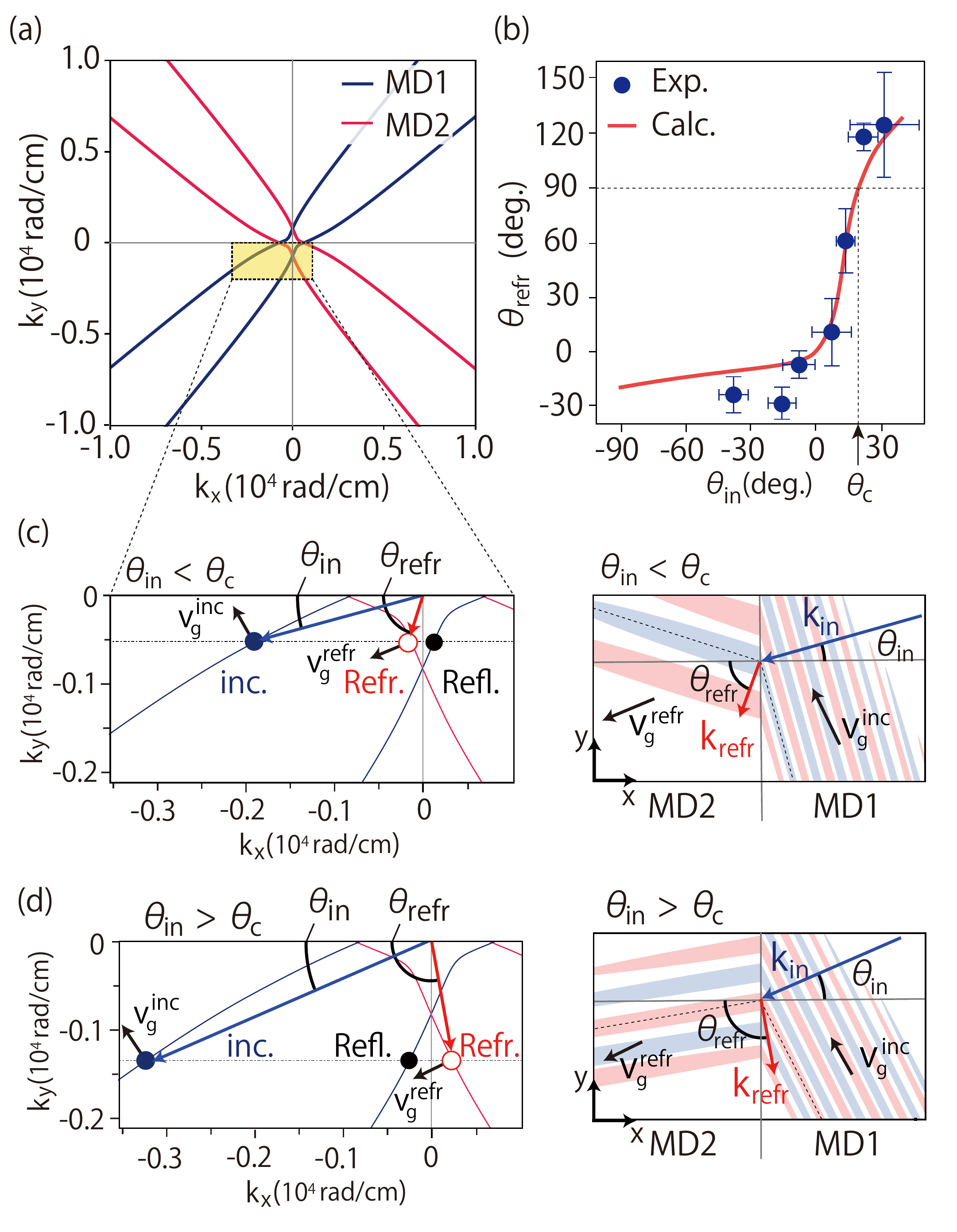}
\caption{\label{Fig4}(a) The red and blue lines represent the spin-wave dispersion in MD1 and MD2, respectively, at the frequency of 1.0 GHz. (b). The experimentally obtained refracted angle as a function of incident angle is plotted by filled circles. Solid line shows the relation between refraction angle and incident angle with the model based on Eq.~\ref{eq:disp}. (c) Relation between incident, refracted and reflected wave in wavenumber space and real space for $\theta_\mathrm{in} < \theta_\mathrm{c}$, (c) Relation between incident, refracted and reflected wave in wavenumber space and real space for $\theta_\mathrm{in} > \theta_\mathrm{c}$.}
\end{figure}

The conservation of $k_y$ leads a reflection and refraction rule for spin waves similar to the Snell's law in geometrical optics.
This is modeled as the following.
First, we write the dispersion relation of spin waves by\cite{1}
\begin{equation}
\label{eq:disp}
\frac{\omega}{\mu_0\gamma} = \sqrt{\left(H + M - \frac{Mkd}{2}\right)\left(H + M - \frac{Mkd}{2}\sin^2\theta_\mathrm{\bf k}\right)},
\end{equation}
where $H$, $M$, and $d$, are the external magnetic field, saturation magnetization, sample thickness, respectively.
The dispersion relation in MD1 is shown in Fig.~\ref{Fig4}(a) by blue solid lines.
Since the orientation of the magnetization in MD2 is rotated by 90 degrees compared to MD1, the dispersion relation in MD2 also rotates by 90 degrees compared to that in MD1 as shown in Fig.~\ref{Fig4}(a) by red solid lines.

The experimentally determined relation between the incident angle ($\theta_\mathrm{in}$) and refraction angle ($\theta_\mathrm{refr}$) at the DW is compered with our model calculation in Fig.~\ref{Fig4}(b).
$\theta_\mathrm{refr}$ increases rapidly for $\theta_\mathrm{in}$ in the region satisfying $\theta_\mathrm{in} > 0$, while it slowly changes in $\theta_\mathrm{in} < 0$.
The $k_y$ and frequency conservation before and after the incidence onto a DW leads a solid red line in Fig.~\ref{Fig4}(b), which shows agreement with the experimental data.
Above the angle $\theta_\mathrm{c} ( = 20.4$ degrees), $\theta_\mathrm{refr}$ surpasses 90 degrees.
This means that the refracted waves propagate towards the inverse direction compared to that in ordinal refraction between materials with positive refraction indices, showing the negative refraction. 

The negative refraction of spin waves originates from the anisotropy of the dispersion relation and consequent difference in the direction of wavevector and group velocity.
The relation between wavevectors of incident and refracted waves is schematically illustrated in Figs.~\ref{Fig4}(c) and (d).
In the case of $\theta_\mathrm{in} < \theta_\mathrm{c}$, the refracted wave has negative $k_x$ and $k_y$ as seen in the left panel of the Fig.~\ref{Fig4}(c).
The incident and refracted waves are chosen to propagates towards left, considering the group velocity shown by a black arrows.
In this case, both $\theta_\mathrm{in}$ and $\theta_\mathrm{refr}$ are less than 90 degrees, therefore, negative refraction does not occur as shown in the right panel of the Fig.~\ref{Fig4}(c).

In the case of  $\theta_\mathrm{in} > \theta_\mathrm{c}$, as shown in the left panel of Fig.~\ref{Fig4}(d), the refracted wave goes across the $k_y$ axis and carries positive $k_x$. 
At this point, the $\theta_\mathrm{refr}$ surpasses 90 degrees.
Although the wavevector of refracted spin waves directs towards MD1, the refracted wave propagates into MD2, because the group velocity of refracted waves still direct towards left.  
Therefore, the negative refraction of spin waves emerges in this case.

Note that the spin wave ray, which represents the energy flow due to spin waves, shows negative refraction in both cases because refracted spin waves has opposite sign of $y$ component of the group velocity compared to the incident spin waves.
The case of $\theta_\mathrm{in} > \theta_\mathrm{c}$ is special in that both wavevector and group velocity demonstrates negative refraction. 
In the region satisfying $\theta_\mathrm{in} < 0$, negative refraction is not realized, because the refracted state is limited in the region with negative $k_x$.
Let us also note that our model does not consider the contribution of MEC at the DW.
Their negligible contributions in the reflection and refraction of spin waves are implied by the agreement of our model with the experimental data.
This is reasonable since the energy scale of MEC is small compared to the energy of magnetostatic spin waves. 

In summary, we observed Snell's law and negative refraction of spin waves at a 90-degree magnetic domain wall in a magnetic garnet film. 
We observed propagation of spin waves crossing a 90-degree domain wall.
The relation between refraction and incident angle is modeled by considering the anisotropic dispersion relation of spin waves in magnetosatatic regime and the wavenumber conservation of spin waves along the DW.
Our observation leads a novel way for the manipulation of the wavevector of spin waves by using a DW.

\begin{acknowledgments}
This work was financially supported by JST-ERATO Grant Number JPMJER1402, JST ERATO Grant Number JPMJER1402, Japan, Grant-in-Aid for Scientific Research on Innovative Area "Nano Spin Conversion Science" (JP26103005) from JSPS KAKENHI, Japan, and World Premier International Research Center Initiative (WPI), Japan.
T.H. is supported by JSPS through a research fellowship for young scientists (No. 18J21004) and acknowledges the support from GP-Spin at Tohoku University.
\end{acknowledgments}

\end{document}